\documentclass[aps,prd,twocolumn,floatfix,longbibliography,nofootinbib]{revtex4-2} 

\usepackage{graphicx}
\usepackage{amsmath,amssymb}
\usepackage{xcolor}
\usepackage[colorlinks=true,linkcolor=blue,citecolor=blue,urlcolor=blue]{hyperref}

\newcommand{\sNN}{\sqrt{s_{\rm NN}}}
\newcommand{\Nch}{\langle N_{\rm ch}\rangle}
\newcommand{\vtwo}[1]{v_{#1}\{2,|\Delta\eta|>2\}}
\newcommand{\ctwo}[1]{c_{#1}\{2\}}
\newcommand{\SC}{\mathrm{SC}}
\newcommand{\NSC}{\mathrm{NSC}}
\newcommand{\czero}{c_{0}}

\begin{document}

\title{Untangling flow-harmonic correlations in OO and Pb--Pb collisions \\ at the LHC using event-shape engineering}

\author{Suraj Prasad}
\email[]{suraj.prasad@cern.ch}
\affiliation{HUN-REN Wigner Research Centre for Physics,
29--33 Konkoly-Thege Mikl\'os Str., H-1121 Budapest, Hungary}
\date{\today}

\begin{abstract}
Symmetric cumulants $\mathrm{SC}(m,n)$ measure the correlation between the event-by-event fluctuations of the flow harmonics $v_m$ and $v_n$, which, for higher harmonics, is influenced by the nonlinear response to the elliptic ($v_2$) and triangular ($v_3$) flow. Event-shape engineering, which selects events by the reduced flow vector $q_n$, is expected to control this contribution. However, its power is limited by the small forward multiplicity in light-ion collisions. In this work, we study the effect of the $q_2$ and $q_3$ selections on the symmetric cumulants in OO and Pb--Pb collisions at $\sqrt{s_{\rm NN}}=5.36$~TeV using EPOS4 with an ATLAS-like acceptance. Using the law of total covariance, the symmetric cumulant is separated into a selected part, which follows the differences between the $q_n$ classes, and a residual part within them. It is observed that in OO collisions the correlations of $v_2$ with $v_3$ and $v_4$ remain almost entirely within the $q_2$ classes, unlike in Pb--Pb collisions. Nevertheless, events with the smallest $q_2$ in central OO collisions show weaker correlations of $v_2$ with $v_3$ and $v_4$, which is attributed to $v_4$ approaching its linear response. In Pb--Pb collisions, the $q_2$ and $q_3$ selections act separately on the two factors of the nonlinear coupling of $v_5$, which establishes a way to isolate them experimentally.
\end{abstract}

\maketitle

\section{Introduction}
\label{sec:intro}

Heavy-ion collisions at the Relativistic Heavy Ion Collider (RHIC) and at the Large Hadron Collider (LHC) produce a hot and dense medium of deconfined quarks and gluons, known as the quark--gluon plasma (QGP)~\cite{STAR:2005gfr,PHENIX:2004vcz,ALICE:2022wpn}, whose formation at high temperature is predicted by lattice QCD~\cite{HotQCD:2014kol}. Since the QGP is short-lived, it cannot be observed directly, and its formation is inferred from the signatures it leaves on the final-state hadrons, such as strangeness enhancement~\cite{Rafelski:1982pu}, jet quenching~\cite{ALICE:2010yje}, quarkonium suppression~\cite{Matsui:1986dk} and collective flow~\cite{Voloshin:2008dg,Heinz:2013th}. Among these, the collective flow probes the expansion of the medium as a whole. Its comparison with hydrodynamic calculations shows that the QGP behaves as a nearly perfect fluid, with a ratio of shear viscosity to entropy density close to the conjectured lower bound of $1/4\pi$~\cite{Kovtun:2004de,Heinz:2013th}.

The collective flow arises because the pressure gradients of the produced medium convert the spatial anisotropy of the overlap region into a momentum-space anisotropy of the final-state particles~\cite{Voloshin:2008dg,Heinz:2013th}. This anisotropy is quantified by the Fourier coefficients $v_n$ of the azimuthal distribution of the particles,
\begin{equation}
\frac{dN}{d\varphi} \propto 1 + 2\sum_{n=1}^{\infty} v_n \cos[n(\varphi-\Psi_n)],
\label{eq:fourier}
\end{equation}
where $\varphi$ is the azimuthal angle of a particle and $\Psi_n$ is the $n$th-order symmetry-plane angle. The coefficients $v_2$ and $v_3$ are known as elliptic and triangular flow. The initial anisotropy, characterized by the eccentricities $\varepsilon_n$, fluctuates from event to event because the positions of the nucleons and of their constituents fluctuate~\cite{Alver:2010gr,Luzum:2011mm}. Hydrodynamic calculations show that $v_2$ and $v_3$ respond nearly linearly to $\varepsilon_2$ and $\varepsilon_3$~\cite{Niemi:2012aj,Gardim:2011xv}, whereas $v_4$ and $v_5$ have two contributions. In terms of the complex flow vectors $V_n=v_ne^{in\Psi_n}$ and eccentricities $\mathcal{E}_n=\varepsilon_ne^{in\Phi_n}$, where $\Phi_n$ is the participant-plane angle,
\begin{align}
V_4 &= \kappa_4\mathcal{E}_4 + \chi_{422}V_2^2, \label{eq:v4}\\
V_5 &= \kappa_5\mathcal{E}_5 + \chi_{523}V_2V_3, \label{eq:v5}
\end{align}
i.e., a linear response to $\varepsilon_4$ and $\varepsilon_5$ with the coefficients $\kappa_n$, and a nonlinear response to $v_2^2$ and to $v_2v_3$ with the coefficients $\chi_{422}$ and $\chi_{523}$~\cite{Gardim:2011xv,Teaney:2012ke,Yan:2015jma,Qian:2016fpi}. The ALICE Collaboration has shown that the nonlinear contributions dominate $v_4$ and $v_5$ in noncentral Pb--Pb collisions~\cite{ALICE:2017fcd,ALICE:2020sup}.

The correlation between the event-by-event fluctuations of two flow harmonics, $v_m$ and $v_n$, is measured with the symmetric cumulant~\cite{Bilandzic:2013kga},
\begin{equation}
\SC(m,n) = \langle v_m^2v_n^2\rangle - \langle v_m^2\rangle\langle v_n^2\rangle,
\label{eq:scdef}
\end{equation}
where the brackets denote the average over events, and with its normalized form $\NSC(m,n)$, which divides $\SC(m,n)$ by $\langle v_m^2\rangle\langle v_n^2\rangle$. In Pb--Pb collisions, $\NSC(3,2)$ is negative, i.e., events with a larger $v_2$ tend to have a smaller $v_3$, which is attributed to the anticorrelation between $\varepsilon_2$ and $\varepsilon_3$ in the initial geometry. In contrast, $\NSC(4,2)$ is positive and is dominated by the nonlinear response of $v_4$ to $v_2^2$~\cite{ALICE:2016kpq,ALICE:2017kwu,ALICE:2021adw}. The symmetric cumulants have also been measured in pp, p--Pb and Xe--Xe collisions~\cite{ALICE:2019zfl,CMS:2017kcs,ATLAS:2018ngv}. It is important to note that in small systems, few-particle correlations from jets and resonance decays, known as nonflow, contribute strongly to the four-particle correlators, and their removal requires subevent methods or explicit subtraction~\cite{Jia:2017hbm,Huo:2017nms,ATLAS:2017rtr,Wang:2026hak,Mallick:2026miz}.

A complementary way to study the coupling of flow harmonics is event-shape engineering, where events of the same centrality are classified according to the magnitude of the reduced flow vector $q_n$ measured in a rapidity window separated from the one used for $v_n$~\cite{Schukraft:2012ah,ALICE:2015lib}. Since $q_n$ follows the event-by-event value of $v_n$, the $q_2$ selection changes the average $v_2$ at fixed centrality, and the response of the other harmonics reveals how they are coupled to $v_2$. Using this method in Pb--Pb collisions, the ATLAS Collaboration found that $v_3$ decreases when $v_2$ is increased by the $q_2$ selection, whereas $v_4$ and $v_5$ increase with $v_2$ and $v_5$ also increases with $v_3$~\cite{ATLAS:2015qwl}, as expected from the sum of a linear and a nonlinear term in hydrodynamics. The ALICE Collaboration has recently observed the same correlations of $v_3$ and $v_4$ with $v_2$ in Pb--Pb and Xe--Xe collisions and has separated the linear and nonlinear parts of $v_4$ in the $q_2$ classes~\cite{ALICE:2026ykw}. As predicted in Ref.~\cite{Magdy:2020bhd} with the multiphase transport model~\cite{Lin:2004en}, the $q_2$ selection changes the nonlinear part strongly, while the linear part and the coupling coefficient $\chi_{422}$ depend little on it, except in the most central collisions. Event-shape selections, also with other observables such as transverse spherocity~\cite{Prasad:2025yfj}, have been used to study the correlation between $v_n$ and the mean transverse momentum~\cite{Magdy:2021ocp}, the initial and final-state anisotropies~\cite{Prasad:2022zbr,Prasad:2025ezg}, global properties of the medium~\cite{Prasad:2021bdq} and symmetry plane correlations~\cite{Tripathy:2025npe}.

In 2025, the LHC provided OO and Ne--Ne collisions at $\sNN=5.36$~TeV. These collisions produce charged-particle multiplicities between those of pp and Pb--Pb collisions, which makes them a bridge between small and large systems~\cite{Nagle:2018nvi,Grosse-Oetringhaus:2024bwr,Brewer:2021kiv,Huss:2020dwe}. The ALICE, ATLAS and CMS collaborations have measured $v_2$ and $v_3$ in OO and Ne--Ne collisions and found that the difference between the two systems follows the different shapes of the $^{16}$O and $^{20}$Ne nuclei~\cite{ALICE:2025luc,ATLAS:2025nnt,CMS:2025tga}, in agreement with hydrodynamic calculations~\cite{Giacalone:2024luz,MenonKavumpadikkalRadhakrishnan:2026tfn}. Multiparticle correlations between harmonics, such as the symmetric cumulants and the symmetry plane correlations, have been proposed as further probes of the nuclear structure and of the medium response in these systems~\cite{Behera:2023nwj,Prasad:2026osg,Mallick:2026miz}.

The event-shape measurements compare the response of the flow harmonics with the symmetric cumulants only qualitatively~\cite{ALICE:2026ykw}. To our knowledge, the symmetric cumulants themselves have not been studied in $q_n$ classes, either in Pb--Pb or in light-ion collisions. It is thus not known how much of the correlation between $v_m$ and $v_n$ fluctuations is carried by the variation of $v_2$ and $v_3$ that the $q_2$ and $q_3$ selections control, and how much remains within the selected $q_n$ classes. This question is particularly important in light-ion collisions, where the multiplicity in the forward window used to measure $q_n$ is about an order of magnitude smaller than in Pb--Pb collisions. The measured $q_n$ then follows the underlying $v_n$ less closely, and the selection may not reduce the lower-order harmonics enough to change the correlations between them.

In this paper, we study the effect of the $q_2$ and $q_3$ selections on the two-particle flow coefficients $v_n\{2\}$ ($n=2$--5) and on the symmetric cumulants $\SC(3,2)$, $\SC(4,2)$, $\SC(4,3)$, $\SC(5,2)$, $\SC(5,3)$ and $\SC(5,4)$ and their normalized forms in OO and Pb--Pb collisions at $\sNN=5.36$~TeV using the EPOS4 model~\cite{Werner:2023zvo,Werner:2023jps}. The pseudorapidity acceptance, the centrality estimator and the $q_n$ definition follow those of the ATLAS event-shape analysis~\cite{ATLAS:2015qwl}, and the transverse-momentum range is $0.2<p_{\rm T}<5$~GeV/$c$. We separate the symmetric cumulant of $q_n$-integrated events into a selected part, $\SC^{\rm sel}$, which follows the differences between the $q_n$ classes, and a residual part, $\SC^{\rm res}$, which remains within them. This shows directly how much of each correlation the selection carries. The dependence on the $q_n$ acceptance is checked with wider $q_n$ windows.

The paper is organized as follows. Section~\ref{sec:method} describes the event samples, the $q_n$ selections, the observables and the nonflow subtraction. Section~\ref{sec:results} presents the results for $v_n\{2\}$ and for the symmetric cumulants. Section~\ref{sec:summary} summarizes the findings. The comparison of EPOS4 with measurements, the dependence on the nonflow suppression and the dependence on the $q_n$ window are given in Appendices~\ref{app:data}, \ref{app:nonflow} and~\ref{app:window}.

\section{Event generation and analysis method}
\label{sec:method}

\subsection{Event generation}

EPOS4 is a general-purpose event generator that treats pp, p--A and A--A collisions within the same framework~\cite{Werner:2023zvo,Werner:2023jps}. The primary interactions are parallel parton--parton scatterings that respect energy conservation and include saturation. The resulting strings are cut into segments, which are separated into a core and a corona according to their energy loss. The core is evolved hydrodynamically and is hadronized microcanonically, whereas the corona segments hadronize as strings. The core fraction grows with the multiplicity, so that the collective response increases continuously from pp to Pb--Pb collisions. EPOS4 has recently been compared with the multiplicity dependence of soft observables, including $v_n\{2\}$, in Pb--Pb, Xe--Xe and OO collisions~\cite{Malik:2026wnx}.

The samples used in this work were generated with EPOS4 version 4.0.3 in its full configuration: the core--corona procedure is activated, the core is evolved hydrodynamically with the standard EPOS equation of state, a nonzero string formation time is used, and the hadronic phase is described by the UrQMD hadronic cascade~\cite{Bass:1998ca,Bleicher:1999xi}. The events are minimum bias, without any trigger or kinematic selection. The decays of $\pi^0$, K$^0_{\rm S}$ and of the weakly decaying strange baryons are switched off, while the other strongly and electromagnetically decaying hadrons are decayed. The generator uses a spherical Woods--Saxon density with radius $R=1.12A^{1/3}-0.86A^{-1/3}$~fm and diffuseness $a=0.54$~fm, i.e., $R\approx2.48$~fm for $^{16}$O and $R\approx6.49$~fm for $^{208}$Pb. The results shown in this paper are based on about $10^8$ OO and $5\times10^6$ Pb--Pb events.

\subsection{Centrality and $q_n$ classes}

The analysis uses charged primary particles with $|\eta|<2.5$ and $0.2<p_{\rm T}<5$~GeV/$c$ for the flow observables. The pseudorapidity range is that of the ATLAS analysis of Ref.~\cite{ATLAS:2015qwl}, which used $p_{\rm T}>0.5$~GeV/$c$. The centrality is defined from the sum of the transverse energy, $\Sigma E_{\rm T}$, of charged primary particles in the acceptance of the ATLAS forward calorimeters, $3.2<|\eta|<4.9$. For Pb--Pb collisions, the centrality classes are 10\% wide, from 0--10\% to 70--80\%. For OO collisions, the centrality classes are 0--10\%, 10--20\%, 20--30\%, 30--40\%, 40--70\% and 70--100\%, i.e., they are wider in peripheral collisions, where the multiplicity is small and the statistical uncertainties of the multiparticle correlators are large.

The reduced flow vector of order $n$ is
\begin{equation}
q_n = \frac{\left|\sum_j E_{{\rm T},j}\, e^{in\varphi_j}\right|}{\sum_j E_{{\rm T},j}}, \quad n=2,3,
\label{eq:qn}
\end{equation}
where the sums run over charged primary particles in $3.3<|\eta|<4.8$ and $E_{\rm T}=E\sin\theta$. This window is separated by at least 0.8 units of pseudorapidity from the tracks used for the flow observables. To study the dependence on the $q_n$ acceptance, $q_n$ is also computed in two wider windows, $3.3<|\eta|<6.3$ and $3.3<|\eta|<7.8$, which extend the default window outward and keep the same separation from the tracks (Appendix~\ref{app:window}). Since $q_n$ changes with multiplicity, a $q_n$ selection over a wide centrality class would partly select the events by their multiplicity instead of their shape. To avoid this, the events of each 1\%-wide centrality interval are sorted by $q_2$ (or $q_3$) and divided into five $q_n$ classes with equal numbers of events. The $q_n$ classes are labeled by the fraction of events with a larger $q_n$, i.e., the 0--20\% $q_n$ class contains the 20\% of events with the largest $q_n$, and the 80--100\% $q_n$ class the 20\% with the smallest $q_n$.\footnote{This is the opposite of the convention of Ref.~\cite{ALICE:2026ykw}, where the 0--10\% $q_n$ class contains the events with the smallest $q_n$.} A $q_n$ class of a wider centrality class, e.g., 0--10\%, combines the same $q_n$ class of all its 1\% intervals, so that each interval contributes the same fraction of its events. In the following, the 0--20\% and 80--100\% $q_n$ classes are called the \emph{highest} and \emph{lowest} $q_n$ classes, and all events of a centrality class, without a $q_n$ selection, are called $q_n$-integrated events.

\subsection{Two-particle and symmetric cumulants}
\label{sec:cumulants}

The two-particle cumulant is computed with the 2-subevent method, with subevent A at $\eta<-1$ and subevent B at $\eta>1$,
\begin{align}
c_n\{2,|\Delta\eta|>2\} &= \langle\langle \cos n(\varphi_1^{\rm A}-\varphi_2^{\rm B})\rangle\rangle, \nonumber\\
\vtwo{n} &= \sqrt{c_n\{2,|\Delta\eta|>2\}},
\label{eq:vn2}
\end{align}
where the double brackets denote the average over pairs and then over events, weighted by the number of pairs. The pseudorapidity gap of two units suppresses short-range nonflow, such as resonance decays and near-side jet fragments. When $c_n\{2,|\Delta\eta|>2\}$ is negative, $\vtwo{n}$ is not defined and that point is not shown.

In terms of multiparticle correlations, the symmetric cumulant of harmonics $m$ and $n$ is~\cite{Bilandzic:2013kga}
\begin{align}
\SC(m,n) &= \langle\langle \cos(m\varphi_1+n\varphi_2-m\varphi_3-n\varphi_4)\rangle\rangle \nonumber\\
&\quad - \langle\langle \cos m(\varphi_1-\varphi_2)\rangle\rangle\langle\langle \cos n(\varphi_1-\varphi_2)\rangle\rangle,
\label{eq:sc}
\end{align}
which equals Eq.~\eqref{eq:scdef} in the absence of nonflow. A positive (negative) $\SC(m,n)$ means that $v_m^2$ and $v_n^2$ are correlated (anticorrelated) from event to event. We study $\SC(3,2)$, $\SC(4,2)$, $\SC(4,3)$, $\SC(5,2)$, $\SC(5,3)$ and $\SC(5,4)$. Through the nonlinear terms of Eqs.~\eqref{eq:v4} and~\eqref{eq:v5}, $\SC(5,2)$ and $\SC(5,3)$ probe the nonlinear part of $v_5$, while $\SC(4,3)$ and $\SC(5,4)$ connect $v_4$ to $v_3$ and $v_5$ to $v_4$ mainly through their common dependence on $v_2$. The normalized symmetric cumulant,
\begin{equation}
\NSC(m,n) = \frac{\SC(m,n)}{\ctwo{m}\,\ctwo{n}},
\label{eq:nsc}
\end{equation}
removes the dependence on the magnitudes of $v_m$ and $v_n$. This is important for the event-shape selection, which changes these magnitudes directly. For example, in the highest $q_2$ class of the 0--10\% OO centrality class, $\SC(4,2)$ is a factor of 1.8 larger than for $q_n$-integrated events, while $\NSC(4,2)$ changes by only 5\%. For this reason, the strength of the correlations in the $q_n$ classes is discussed with $\NSC(m,n)$, whose significance is nearly the same as that of $\SC(m,n)$, since $\ctwo{n}$ is much more precise than the symmetric cumulants. Here $\ctwo{n}=\langle\langle \cos n(\varphi_1-\varphi_2)\rangle\rangle$ is computed with all particles in $|\eta|<2.5$, which is the same particle set as the four-particle correlator. All multiparticle correlators are calculated with the $Q$-cumulant method~\cite{Bilandzic:2010jr,Bilandzic:2013kga}. The symmetric cumulants are computed in bins of unit multiplicity $M$ of the tracks and then averaged with the weight $W_4(M)=M(M-1)(M-2)(M-3)$, which reduces the effect of multiplicity fluctuations within a centrality or $q_n$ class. The statistical uncertainties are obtained with the delete-one jackknife method using 10 subsamples. The significance of a difference between a $q_n$ class and $q_n$-integrated events is computed with their uncertainties added in quadrature, which neglects their positive correlation and therefore underestimates the significance.

\subsection{Nonflow subtraction}

The four-particle correlator in Eq.~\eqref{eq:sc} still contains nonflow, which grows towards low multiplicity as $1/M^3$ in the independent-source picture~\cite{Borghini:2001vi,Wang:2026hak}. Following Refs.~\cite{Wang:2026hak,Mallick:2026miz}, nonflow is subtracted with a template from the low-multiplicity (LM) class, which is the 70--100\% centrality class of $q_n$-integrated events,
\begin{equation}
\SC^{\rm sub}(m,n) = \SC(m,n) - f\, \SC_{\rm LM}(m,n).
\label{eq:sub}
\end{equation}
The scale factor follows the multiplicity dependence of nonflow ($\czero$ scaling), $f=(N_{4,{\rm LM}}/N_4)^3$, where $1/N_4^3=\sum_M W_4(M) M^{-3}/\sum_M W_4(M)$ is the effective multiplicity of the centrality or $q_n$ class. The scale factor is computed in every centrality and $q_n$ class, and the denominator of Eq.~\eqref{eq:nsc} is not subtracted. In the 0--40\% OO centrality classes, the subtraction changes $\SC(3,2)$ by a few percent at most, while the fraction of $\SC(4,2)$ that it removes grows from 2\% in the 0--10\% centrality class to 37\% in the 30--40\% centrality class. In the 40--70\% centrality class, it removes most of $\SC(4,2)$, so the results there depend on how well the LM template describes the nonflow. The symmetric cumulants are shown up to 70\% centrality in OO collisions and up to 60\% in Pb--Pb collisions. For OO collisions, however, we restrict the discussion to the centrality classes where $\SC(m,n)$ of $q_n$-integrated events has a significance of about three standard deviations or more and the subtraction is a moderate correction, i.e., 0--40\% for $(m,n)=(3,2)$, 0--30\% for $(4,2)$, 0--20\% for $(4,3)$ and $(5,2)$, and 0--10\% for $(5,3)$. The 70--100\% OO centrality class is the LM class itself, so its subtracted symmetric cumulants vanish by construction, and it is used only for $v_n\{2\}$.

\subsection{Selected and residual correlation}
\label{sec:decomp}

A $q_n$ selection divides the events of a centrality class into $q_n$ classes $k$, each with a fraction $w_k$ of the events; here, these are the five $q_n$ classes of Sec.~\ref{sec:method}, so that $w_k=0.2$. Since $\SC(m,n)$ is the covariance of $v_m^2$ and $v_n^2$, Eq.~\eqref{eq:scdef}, it can be split into two parts with the law of total covariance. The first part, which we call the residual part, $\SC^{\rm res}(m,n)$, is the correlation that remains among events of similar $q_n$: $\SC(m,n)$ is computed in each $q_n$ class and averaged over the $q_n$ classes. The second part, the selected part $\SC^{\rm sel}(m,n)$, is the correlation between the averages of the $q_n$ classes: it is nonzero only if the $q_n$ classes with a larger $\langle v_m^2\rangle$ also have a larger (or smaller) $\langle v_n^2\rangle$. The two parts add up to $\SC(m,n)$ of $q_n$-integrated events,
\begin{align}
\SC(m,n) =&\ \underbrace{\sum_k w_k\,\SC_k(m,n)}_{\SC^{\rm res}(m,n)} \nonumber\\
&+ \underbrace{\sum_k w_k\,\big(\ctwo{m}_k-\overline{c}_m\big)\big(\ctwo{n}_k-\overline{c}_n\big)}_{\SC^{\rm sel}(m,n)},
\label{eq:decomp}
\end{align}
where $\SC_k$ and $\ctwo{n}_k$ are measured in $q_n$ class $k$ and $\overline{c}_n=\sum_k w_k\ctwo{n}_k$. In the statistics literature, the two parts are known as the within-class and between-class terms~\cite{Ross:2002prob}.

Two limiting cases illustrate the meaning of the two parts. If $q_2$ measured $v_2$ exactly and the $q_2$ classes were very narrow, all events of a $q_2$ class would have the same $v_2$, and $\SC_k(m,2)$ would vanish in every $q_2$ class. The whole correlation of $v_m$ with $v_2$ would then appear as a difference of $\langle v_m^2\rangle$ and $\langle v_2^2\rangle$ between the $q_2$ classes, i.e., it would be in $\SC^{\rm sel}$. In the opposite case, $q_2$ carries no information on $v_2$, all $q_2$ classes have the same $\langle v_m^2\rangle$ and $\langle v_2^2\rangle$, and the whole correlation is in $\SC^{\rm res}$. A real selection lies between these two cases, since $q_n$ is measured with a finite number of particles at forward rapidity and follows $v_n$ at midrapidity only partly. Therefore, $\SC^{\rm sel}$ measures how much of the correlation the $q_n$ selection controls, and $\SC^{\rm res}$ is what the selection cannot remove.

The sign of $\SC^{\rm sel}$ follows from how the two harmonics respond to the selection. For example, if the highest $q_2$ class has a larger $v_2$ and a smaller $v_3$ than average, and the lowest $q_2$ class the opposite, as observed in Pb--Pb collisions (Sec.~\ref{sec:vn}), the products in Eq.~\eqref{eq:decomp} are negative, and $\SC^{\rm sel}(3,2)$ is negative, like $\SC(3,2)$ itself. If, instead, the selection changes only one harmonic of the pair, $\SC^{\rm sel}$ is small, irrespective of how strongly that harmonic changes. This is the case for $\SC(4,2)$ with the $q_3$ selection, whose $q_3$ classes differ in $v_3$ but hardly in $v_2$. It is important to note that the two parts need not have the same sign: when $\SC^{\rm sel}(m,n)$ has the sign opposite to that of $\SC(m,n)$, $\SC^{\rm res}(m,n)$ is larger in magnitude than $\SC(m,n)$ itself, as for $\SC(5,3)$ with the $q_2$ selection in Pb--Pb collisions (Sec.~\ref{sec:selected}).

Both parts are divided by $\ctwo{m}\ctwo{n}$ of $q_n$-integrated events, which gives $\NSC^{\rm sel}(m,n)$ and $\NSC^{\rm res}(m,n)$, so that their sum is $\NSC(m,n)$ of $q_n$-integrated events. Note that $\NSC^{\rm res}(m,n)$ is not $\NSC(m,n)$ of one $q_n$ class, since it uses the denominator of $q_n$-integrated events. The $\czero$ subtraction of each $q_n$ class enters $\SC^{\rm res}$, while $\SC^{\rm sel}$ is built from $\ctwo{n}$ of the $q_n$ classes. The size of $\SC^{\rm sel}$ also depends on the number of $q_n$ classes, as finer $q_n$ classes resolve more of the variation of $v_n$ with $q_n$; with ten $q_n$ classes of 10\%, it increases only slightly. Since the symmetric cumulants are averaged over multiplicity bins with the weight $W_4(M)$, Eq.~\eqref{eq:decomp} is not exact. We have verified that the sum of the two parts reproduces $\SC(m,n)$ of $q_n$-integrated events within 2.1 times its statistical uncertainty, and within 4\% of its value where $\SC(m,n)$ is significant, for every system, centrality class, harmonic pair and selection shown here.

\section{Results and discussion}
\label{sec:results}

\subsection{Response of the flow coefficients to the $q_2$ and $q_3$ selections}
\label{sec:vn}

\begin{figure*}
\includegraphics[width=\textwidth]{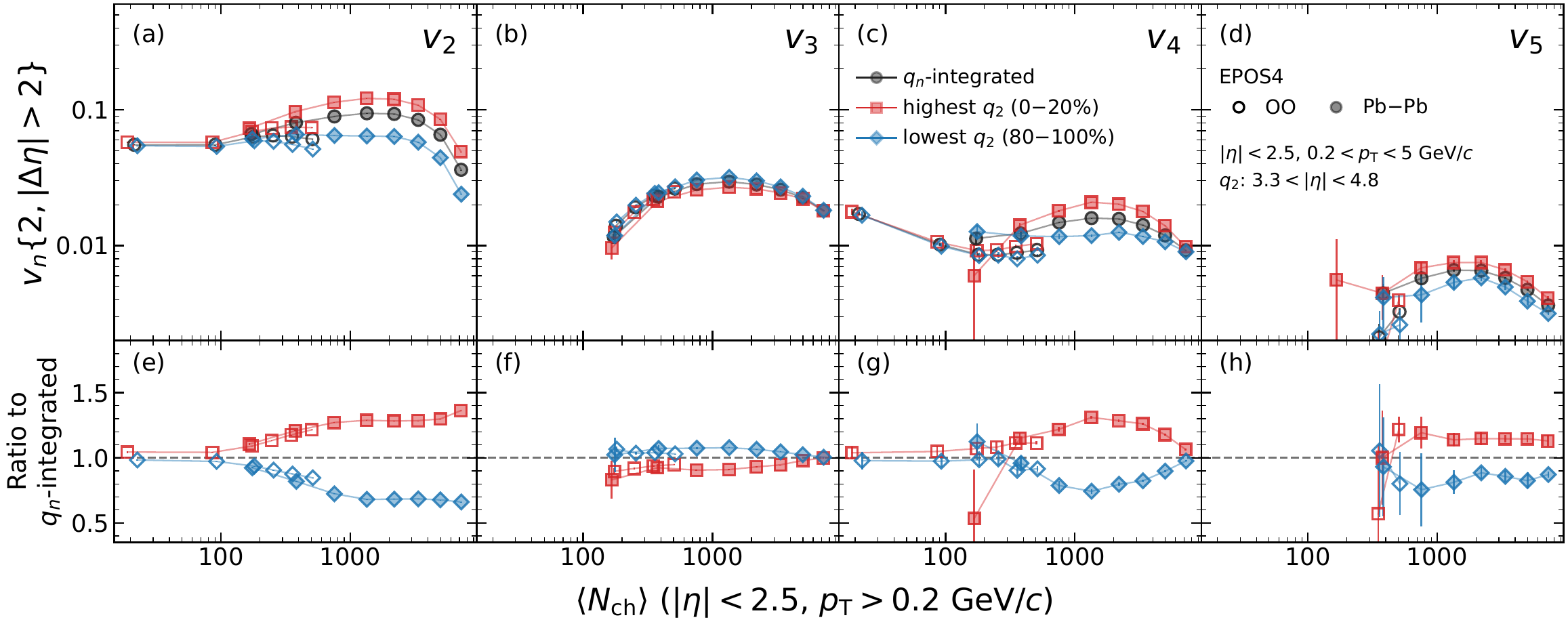}
\caption{Upper panels: $\vtwo{2}$ (a), $\vtwo{3}$ (b), $\vtwo{4}$ (c) and $\vtwo{5}$ (d) as a function of $\Nch$ in OO and Pb--Pb collisions at $\sNN=5.36$~TeV using EPOS4, for $q_n$-integrated events (black) and for the highest (0--20\%, red) and lowest (80--100\%, blue) $q_2$ classes. Each point corresponds to one centrality class and is placed at the $\Nch$ of its own events. Lower panels (e)--(h): ratio of the $q_2$ class to $q_n$-integrated events. Centrality classes with a negative $c_n\{2,|\Delta\eta|>2\}$ are not shown. The uncertainties are statistical.}
\label{fig:vnq2}
\end{figure*}

\begin{figure*}
\includegraphics[width=\textwidth]{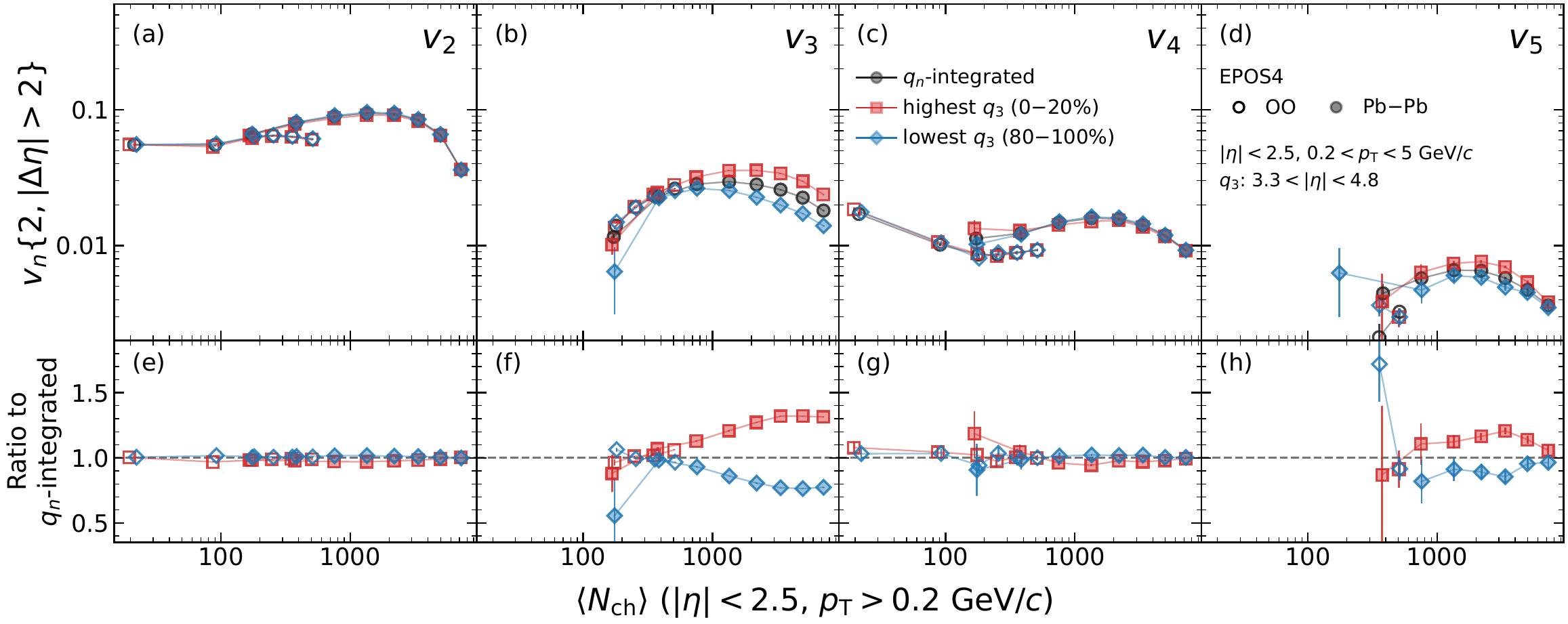}
\caption{Upper panels: $\vtwo{2}$ (a), $\vtwo{3}$ (b), $\vtwo{4}$ (c) and $\vtwo{5}$ (d) as a function of $\Nch$ in OO and Pb--Pb collisions at $\sNN=5.36$~TeV using EPOS4, for $q_n$-integrated events (black) and for the highest (0--20\%, red) and lowest (80--100\%, blue) $q_3$ classes. Each point corresponds to one centrality class and is placed at the $\Nch$ of its own events. Lower panels (e)--(h): ratio of the $q_3$ class to $q_n$-integrated events. Centrality classes with a negative $c_n\{2,|\Delta\eta|>2\}$ are not shown. The uncertainties are statistical.}
\label{fig:vnq3}
\end{figure*}

An event-shape selection can change the correlations between flow harmonics only to the extent that it changes the harmonics themselves. We therefore begin with the flow coefficients. Figure~\ref{fig:vnq2} shows $\vtwo{n}$ for $n=2$--5 as a function of $\Nch$ in OO and Pb--Pb collisions, for $q_n$-integrated events and for the highest and lowest $q_2$ classes. For $q_n$-integrated events, $\vtwo{2}$ in OO collisions depends weakly on $\Nch$, whereas in Pb--Pb collisions it increases strongly from central to mid-central collisions, following the elliptic shape of the overlap region, and decreases again in peripheral collisions. The higher harmonics are smaller and depend less on centrality, since they originate mainly from fluctuations of the initial geometry~\cite{Alver:2010gr,Luzum:2011mm,Niemi:2012aj}. It is important to note that the two-unit pseudorapidity gap does not remove the away-side jet correlations, which grow towards low multiplicity~\cite{Jia:2017hbm} and, being peaked at $\Delta\varphi\approx\pi$, contribute with a negative sign to the odd harmonics. As a result, $c_3\{2,|\Delta\eta|>2\}$ is negative in OO collisions below $\Nch\approx100$, and $c_5\{2,|\Delta\eta|>2\}$ is positive only in the 0--10\% and 10--20\% OO centrality classes.

The lower panels of Fig.~\ref{fig:vnq2} show the ratio of $\vtwo{n}$ in the $q_2$ classes to that of $q_n$-integrated events. One can observe that the highest $q_2$ class has a larger $\vtwo{2}$ and the lowest $q_2$ class a smaller one, irrespective of the collision system, which confirms that the forward $q_2$ follows $v_2$ at midrapidity. In OO collisions, the change in the highest $q_2$ class grows with $\Nch$ and reaches about 20\% in the 0--10\% centrality class, although in the peripheral centrality classes it can also contain nonflow correlations between the forward window and midrapidity. At the same $\Nch$, the ratios for OO and peripheral Pb--Pb collisions are of similar size, with a somewhat stronger selection in Pb--Pb collisions at $\Nch\approx370$. This implies that the power of the $q_2$ selection is set mainly by the multiplicity rather than by the collision system. This is because the measured $q_2$ is the sum of the underlying flow vector and of a statistical fluctuation due to the finite number of particles in the forward window, and the latter decreases with multiplicity~\cite{Voloshin:2008dg,Schukraft:2012ah,Jia:2014jca}. In the 0--50\% Pb--Pb centrality classes, the $q_2$ selection increases $\vtwo{2}$ by about 30\% in the highest $q_2$ class and decreases it by about 30\% in the lowest $q_2$ class.

The other harmonics respond to the $q_2$ selection through their coupling to $v_2$. The triangular flow, $\vtwo{3}$, responds with the opposite sign in both systems: it is smaller in the highest $q_2$ class and larger in the lowest $q_2$ class, by up to about 10\%. This anticorrelation between $v_2$ and $v_3$ was observed by ATLAS in Pb--Pb collisions with the same selection~\cite{ATLAS:2015qwl} and originates from the anticorrelation between $\varepsilon_2$ and $\varepsilon_3$ in the initial geometry: an overlap region that is strongly elongated leaves less room for a triangular deformation~\cite{ALICE:2016kpq,Giacalone:2016afq,Zhu:2016kix}. In contrast, $\vtwo{4}$ follows $\vtwo{2}$, and, as for $\vtwo{2}$, its change is larger in Pb--Pb collisions than in OO collisions. In the lowest $q_2$ class of OO collisions, the decrease of $\vtwo{4}$ is seen only in the 0--10\% and 10--20\% centrality classes. This is the expected behavior of the nonlinear response of Eq.~\eqref{eq:v4}, i.e., the $q_2$ selection changes the nonlinear term through $v_2^2$, while it leaves the linear term in $\varepsilon_4$ nearly unchanged~\cite{Magdy:2020bhd,ALICE:2026ykw}. As a result, the lowest $q_2$ class contains events in which $v_4$ is closer to its linear response. The pentagonal flow has a nonlinear term proportional to $V_2V_3$ [Eq.~\eqref{eq:v5}]~\cite{ALICE:2020sup}. In Pb--Pb collisions, $\vtwo{5}$ also follows $\vtwo{2}$, but in mid-central collisions its change is smaller than that of $\vtwo{4}$, since the increase of $v_2$ is partly compensated by the decrease of $v_3$ in the product $v_2v_3$. In OO collisions, $\vtwo{5}$ is defined only in the 0--10\% and 10--20\% centrality classes and has large uncertainties in the $q_2$ classes.

Figure~\ref{fig:vnq3} shows the same quantities for the $q_3$ selection. In Pb--Pb collisions, the $q_3$ selection changes $\vtwo{3}$ by about 20--30\% in the 0--40\% centrality classes. In OO collisions, the change is much weaker: only $+6\%$ and $-3\%$ in the highest and lowest $q_3$ classes of the 0--10\% centrality class, and it decreases further towards peripheral collisions. In the 30--40\% centrality class, the ordering is even reversed, so that the $q_3$ selection does not resolve $v_3$ there, and a nonflow origin of this reversal cannot be excluded. The weaker response, compared with that of $v_2$ to the $q_2$ selection at the same $\Nch$, is because $v_3$ is at least a factor of two smaller than $v_2$, so that the statistical fluctuation of $q_3$ in the forward window dominates over the triangular flow. The $q_3$ selection decreases $\vtwo{2}$ slightly in the highest $q_3$ class, as expected from the anticorrelation of $\varepsilon_2$ and $\varepsilon_3$, and leaves $\vtwo{4}$ nearly unchanged, since the nonlinear term of $v_4$ depends on $v_2^2$ and not on $v_3$. On the other hand, $\vtwo{5}$ of Pb--Pb collisions increases in the highest $q_3$ class, which reflects the nonlinear coupling of $v_5$ to $v_3$. The $q_2$ and $q_3$ selections thus act on the higher harmonics through the same nonlinear couplings that produce the correlations studied below.

\subsection{Correlations of $v_2$ with $v_3$ and $v_4$: $\NSC(3,2)$ and $\NSC(4,2)$}
\label{sec:nsc32}

\begin{figure}
\includegraphics[width=\linewidth]{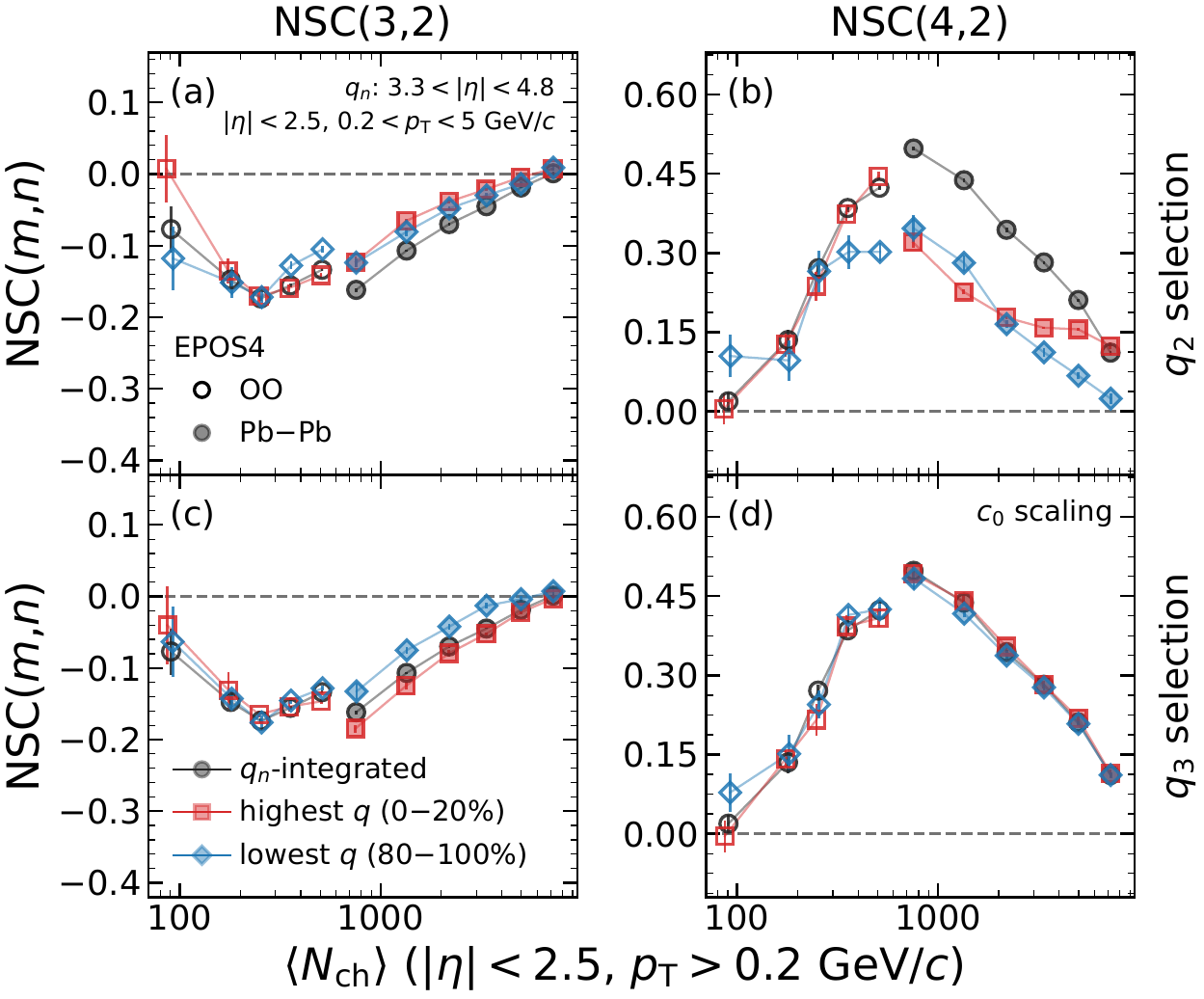}
\caption{$\NSC(3,2)$ (left) and $\NSC(4,2)$ (right) as a function of $\Nch$ in OO and Pb--Pb collisions at $\sNN=5.36$~TeV using EPOS4, for $q_n$-integrated events (black) and for the highest (red) and lowest (blue) classes of the $q_2$ (upper) and $q_3$ (lower) selections, with the $\czero$ nonflow subtraction, in the centrality ranges given in Sec.~\ref{sec:method}. The uncertainties are statistical.}
\label{fig:nsc32}
\end{figure}

\begin{figure*}
\includegraphics[width=0.86\textwidth]{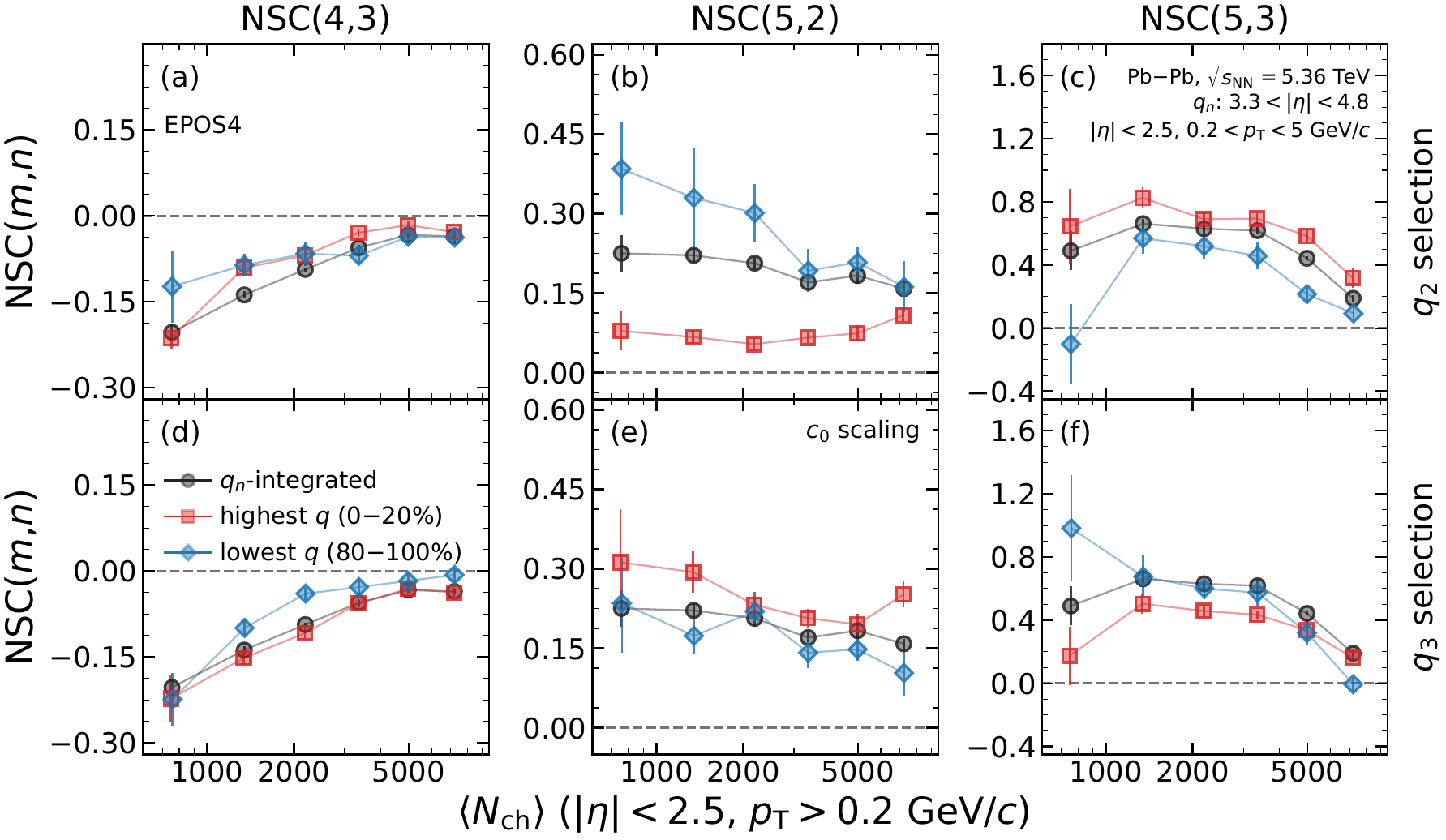}
\caption{$\NSC(4,3)$, $\NSC(5,2)$ and $\NSC(5,3)$ (columns from left to right) as a function of $\Nch$ in Pb--Pb collisions at $\sNN=5.36$~TeV using EPOS4 for $q_n$-integrated events (black) and for the highest (red) and lowest (blue) classes of the $q_2$ (upper) and $q_3$ (lower) selections, with the $\czero$ nonflow subtraction, in the 0--60\% centrality classes. The uncertainties are statistical.}
\label{fig:nscnl}
\end{figure*}

Figures~\ref{fig:nsc32}(a) and (c) show $\NSC(3,2)$ as a function of $\Nch$. For $q_n$-integrated events, $\NSC(3,2)$ is negative in OO and Pb--Pb collisions, which reflects the anticorrelation of $v_2$ and $v_3$ fluctuations inherited from $\varepsilon_2$ and $\varepsilon_3$~\cite{Bilandzic:2013kga,ALICE:2016kpq,Giacalone:2016afq,Zhu:2016kix}. In OO collisions, its magnitude is largest in the 20--30\% centrality class and decreases towards central collisions, whereas in Pb--Pb collisions it decreases steadily from peripheral collisions to about zero in the 0--10\% centrality class. The magnitude of $\NSC(3,2)$ in Pb--Pb collisions is smaller than measured by ALICE at $\sNN=5.02$~TeV (Appendix~\ref{app:data}).

In OO collisions, the $q_2$ selection changes $\NSC(3,2)$ mainly in the lowest $q_2$ class of central collisions. Compared with $q_n$-integrated events, its magnitude is 22\% and 18\% smaller in the 0--10\% and 10--20\% centrality classes, with a significance of 5.6 and 4.6 standard deviations, respectively. The highest $q_2$ class stays close to $q_n$-integrated events. In the 20--30\% and 30--40\% centrality classes, both $q_2$ classes agree with $q_n$-integrated events within their uncertainties, i.e., the reduction disappears as the $q_2$ selection becomes weaker. Events with a small $v_2$ thus show a weaker anticorrelation between $v_2$ and $v_3$, as expected if a part of $v_2$ is driven by the same fluctuations that produce $v_3$. The $q_3$ selection has a smaller effect: only the highest $q_3$ class of the 0--10\% centrality class differs from $q_n$-integrated events, with a slightly larger magnitude. In Pb--Pb collisions, both the highest and the lowest $q_2$ classes give values of $\NSC(3,2)$ closer to zero than $q_n$-integrated events. This is not a contradiction, since $\NSC(3,2)$ of $q_n$-integrated events is not the average of its values in the $q_2$ classes: the $q_2$ selection in Pb--Pb collisions turns a large part of the $v_2$ fluctuations into a difference between the $q_2$ classes, and the anticorrelation of $v_2$ and $v_3$ then appears in $\NSC^{\rm sel}(3,2)$, as a difference of $\langle v_3^2\rangle$ between the $q_2$ classes, rather than in $\NSC^{\rm res}(3,2)$ (Sec.~\ref{sec:selected}). The $q_3$ selection behaves differently: in the 10--60\% Pb--Pb centrality classes, the highest $q_3$ class gives a larger magnitude than $q_n$-integrated events, in the same direction as in OO collisions, and the lowest $q_3$ class gives a smaller one.

Figures~\ref{fig:nsc32}(b) and (d) show $\NSC(4,2)$, which in hydrodynamics is dominated by the nonlinear response of $v_4$ to $v_2^2$~\cite{Giacalone:2016afq,Zhu:2016kix,ALICE:2017kwu}. For $q_n$-integrated events, $\NSC(4,2)$ is positive, increases with $\Nch$ in OO collisions and connects to the values of peripheral Pb--Pb collisions, after which it decreases towards central Pb--Pb collisions, as also reported in Ref.~\cite{Mallick:2026miz} with the multiphase transport model. In OO collisions, the highest $q_2$ class stays close to $q_n$-integrated events. In contrast, the lowest $q_2$ class gives a 29\% smaller value in the 0--10\% centrality class, with a significance of 6.6 standard deviations, and a weaker reduction in the 10--20\% centrality class, while no change is seen in the 20--30\% centrality class. The same reduction of $\NSC(3,2)$ and $\NSC(4,2)$ is found with the 2-subevent method, which suppresses nonflow without a template (Appendix~\ref{app:nonflow}). The reduction of $\NSC(4,2)$ is attributed to the reduction of the nonlinear contribution to $v_4$ in events with a small $v_2$ seen in Fig.~\ref{fig:vnq2}: when $v_2$ is small, $v_4$ is closer to its linear response to $\varepsilon_4$, which is only weakly correlated with $\varepsilon_2$~\cite{Giacalone:2016afq}. In the 10--50\% Pb--Pb centrality classes, both the highest and the lowest $q_2$ classes give a much smaller $\NSC(4,2)$ than $q_n$-integrated events, as for $\NSC(3,2)$. The $q_3$ selection leaves $\NSC(4,2)$ essentially unchanged in both systems.

\begin{figure}
\includegraphics[width=\linewidth]{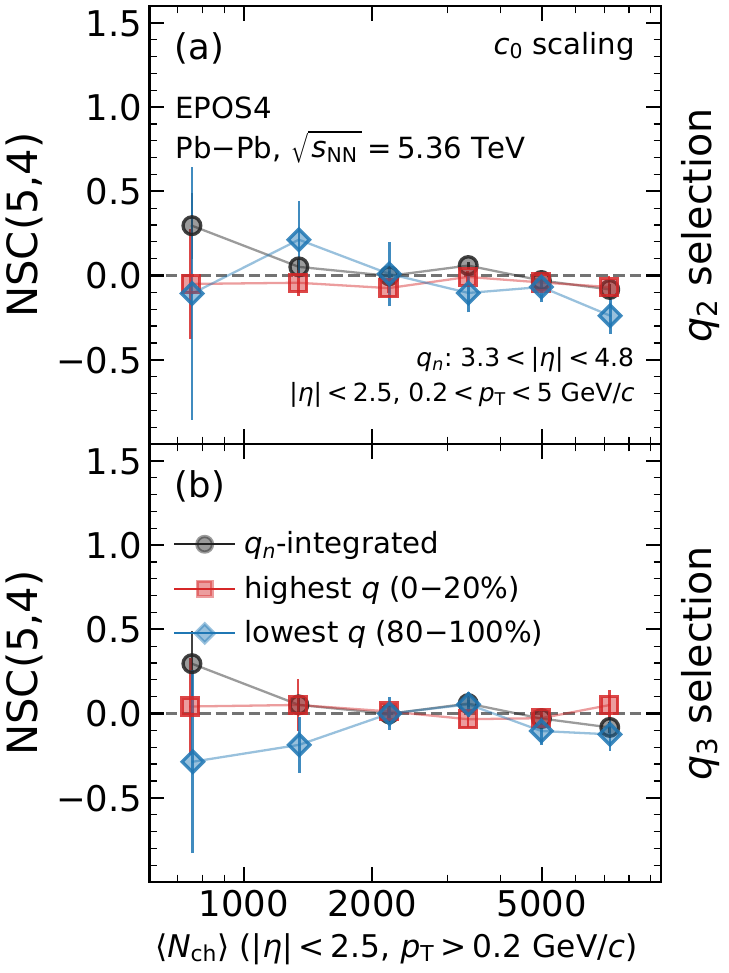}
\caption{Same as Fig.~\ref{fig:nscnl} for $\NSC(5,4)$, for the $q_2$ (a) and $q_3$ (b) selections. The OO results are not shown, since their uncertainties are of order unity.}
\label{fig:nsc54}
\end{figure}

\subsection{Correlations of $v_4$ and $v_5$ with the lower harmonics: $\NSC(4,3)$, $\NSC(5,2)$ and $\NSC(5,3)$}
\label{sec:nonlinear}

Figure~\ref{fig:nscnl} shows $\NSC(4,3)$, $\NSC(5,2)$ and $\NSC(5,3)$ as a function of $\Nch$ in Pb--Pb collisions. For $q_n$-integrated events, $\NSC(4,3)$ is negative, as measured by ALICE~\cite{ALICE:2017kwu,ALICE:2021adw}, and its magnitude grows from central to peripheral collisions. This is because the nonlinear part of $v_4$ follows $v_2^2$, and $v_2$ is anticorrelated with $v_3$~\cite{Zhu:2016kix}. Both $\NSC(5,2)$ and $\NSC(5,3)$ are positive, as also measured~\cite{ALICE:2017kwu}, which is expected from the nonlinear response of $v_5$ to $v_2v_3$~\cite{Zhu:2016kix}.

\begin{figure*}[p]
\includegraphics[width=0.86\textwidth]{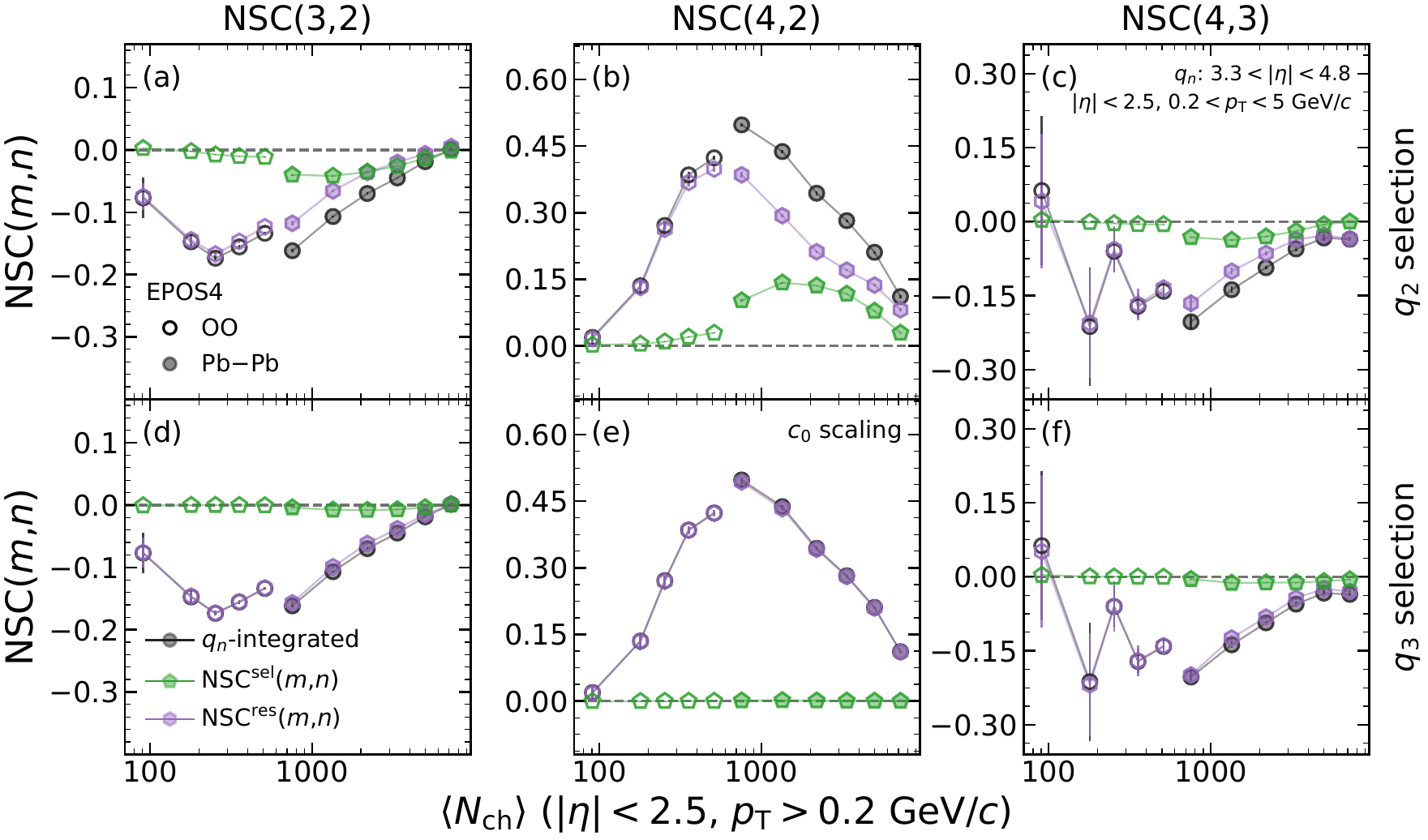}
\caption{$\NSC(3,2)$, $\NSC(4,2)$ and $\NSC(4,3)$ (columns from left to right) of $q_n$-integrated events (black), separated into $\NSC^{\rm sel}(m,n)$ (green) and $\NSC^{\rm res}(m,n)$ (purple) according to Eq.~\eqref{eq:decomp}, as a function of $\Nch$ in OO and Pb--Pb collisions at $\sNN=5.36$~TeV using EPOS4, for the $q_2$ (upper) and $q_3$ (lower) selections, in the centrality ranges given in Sec.~\ref{sec:method}. The $\czero$ nonflow subtraction enters $\NSC^{\rm res}(m,n)$. The uncertainties are statistical.}
\label{fig:decomp1}
\end{figure*}

The response of $\NSC(4,3)$ to the event-shape selection resembles that of $\NSC(3,2)$ and $\NSC(4,2)$. In mid-central collisions, its magnitude is smaller in both the highest and the lowest $q_2$ classes than in $q_n$-integrated events, most clearly in the highest one, since the $q_2$ selection turns a part of the $v_2$ fluctuations, which couple $v_4$ to $v_3$, into a difference between the $q_2$ classes. The $q_3$ selection reduces the magnitude mainly in the lowest $q_3$ class.

The most interesting feature is that the $q_2$ selection changes $\NSC(5,2)$ and $\NSC(5,3)$ in opposite directions. In the highest $q_2$ class of the 10--40\% centrality classes, $\NSC(5,2)$ decreases to between one quarter and two fifths of its value for $q_n$-integrated events, while $\NSC(5,3)$ increases. This increase is small in the individual centrality classes, but it has the same sign in all of the 0--50\% centrality classes, and its weighted mean is 17\% of the value for $q_n$-integrated events, with a significance of 4.1 standard deviations. In the lowest $q_2$ class, $\NSC(5,3)$ decreases, so that the highest $q_2$ class exceeds the lowest one by 48\% of the value for $q_n$-integrated events on average, with a significance of 6.3 standard deviations. This is a direct consequence of the nonlinear term $\chi_{523}V_2V_3$. Within a $q_2$ class, $v_2$ varies little, so that the correlation of $v_5$ with $v_2$ is removed, while the fluctuations of $v_5$ that remain follow those of $v_3$ with a strength proportional to $v_2$. The larger $v_2$ of the highest $q_2$ class thus strengthens the correlation of $v_5$ with $v_3$, and the small $v_2$ of the lowest $q_2$ class weakens it in favor of the linear response of $v_5$ to $\varepsilon_5$. The $q_3$ selection shows the reverse pattern: in the highest $q_3$ class of the 10--50\% centrality classes, $\NSC(5,3)$ decreases to about three quarters of its value for $q_n$-integrated events, while $\NSC(5,2)$ is larger. This is the same mechanism with the roles of $v_2$ and $v_3$ exchanged, i.e., the $q_2$ and $q_3$ selections separate the two factors of the nonlinear coupling.

\begin{figure*}[p]
\includegraphics[width=0.86\textwidth]{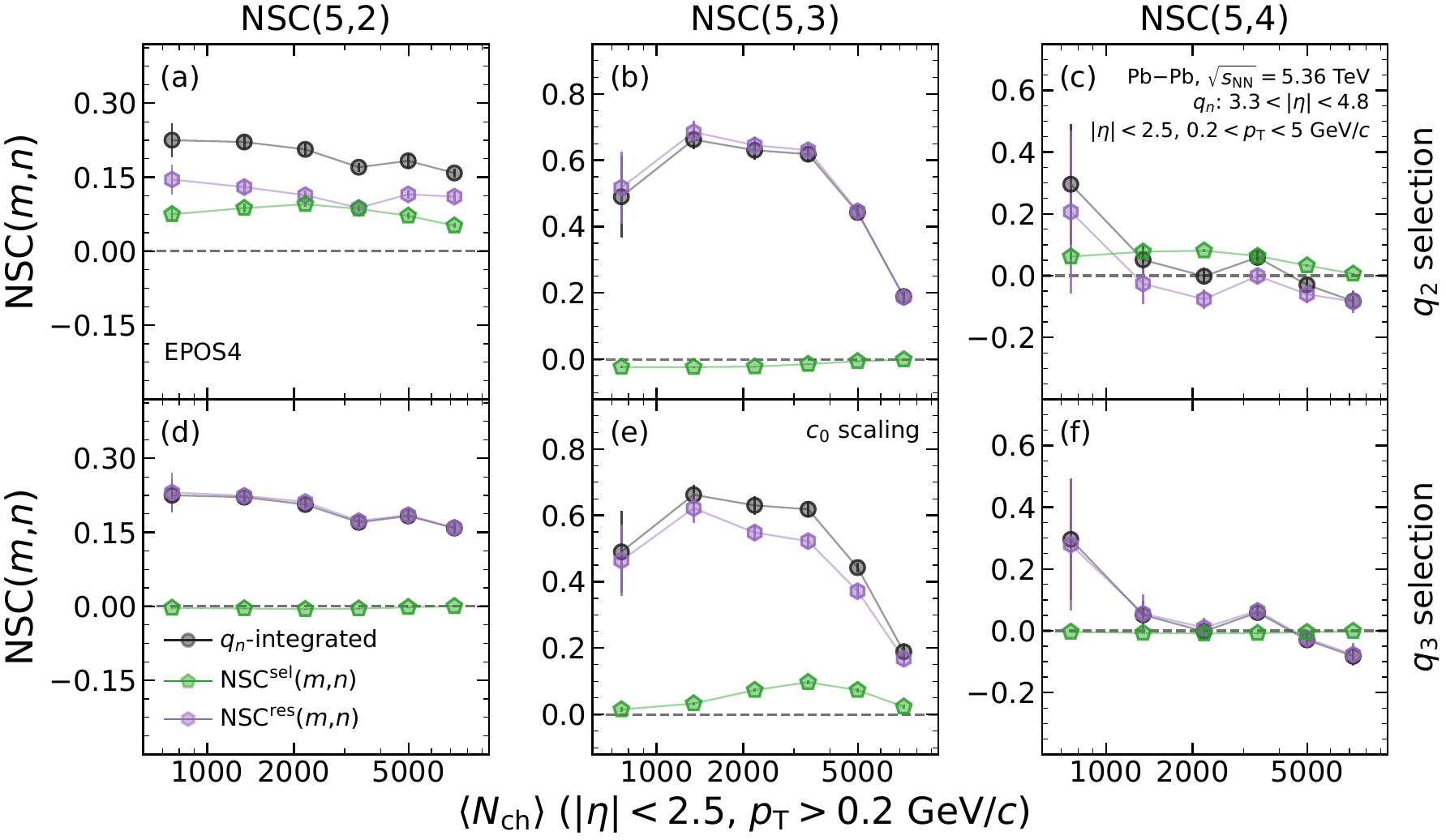}
\caption{$\NSC(5,2)$, $\NSC(5,3)$ and $\NSC(5,4)$ (columns from left to right) of $q_n$-integrated events (black), separated into $\NSC^{\rm sel}(m,n)$ (green) and $\NSC^{\rm res}(m,n)$ (purple) according to Eq.~\eqref{eq:decomp}, as a function of $\Nch$ in Pb--Pb collisions at $\sNN=5.36$~TeV using EPOS4, for the $q_2$ (upper) and $q_3$ (lower) selections, in the 0--60\% centrality classes. The $\czero$ nonflow subtraction enters $\NSC^{\rm res}(m,n)$. The uncertainties are statistical.}
\label{fig:decomp2}
\end{figure*}

In OO collisions, these correlators are significant only in the most central centrality classes (Sec.~\ref{sec:method}), and they are therefore not shown in Fig.~\ref{fig:nscnl}. There, they have the same signs as in Pb--Pb collisions, and the highest $q_2$ class shows the same trends, i.e., a smaller $\NSC(5,2)$ and a larger $\NSC(5,3)$. However, the changes with the $q_2$ and $q_3$ selections are not significant, so that the present OO sample cannot establish them.

\subsection{Correlation between $v_4$ and $v_5$}
\label{sec:nsc54}

Figure~\ref{fig:nsc54} shows $\NSC(5,4)$ in Pb--Pb collisions. Through the nonlinear terms of Eqs.~\eqref{eq:v4} and~\eqref{eq:v5}, $v_4$ and $v_5$ share a dependence on $v_2$, which would correlate them positively, while the anticorrelation of $v_2$ and $v_3$ enters $v_5$ with the opposite sign, so that the two contributions partly cancel. One can observe that $\NSC(5,4)$ of $q_n$-integrated Pb--Pb events is small, with both signs, and it differs from zero by at most three standard deviations in the 0--50\% centrality classes, as do the values in the $q_n$ classes. In OO collisions, $c_4\{2\}c_5\{2\}$ is of the order of $10^{-9}$, which makes the uncertainties of $\NSC(5,4)$ of order unity, and it is therefore not shown. In OO collisions, $\SC(5,4)$ differs from zero by at most 2.1 standard deviations, except in the 10--20\% centrality class, where it is negative with a significance of 5.0 standard deviations. The adjacent centrality classes do not show it, so we do not interpret it as a flow correlation. A measurement of this correlator in OO collisions requires much larger samples than those of the other symmetric cumulants, and the $q_n$ selection does not change this.

\subsection{How much of the correlation is carried by the $q_n$ selection}
\label{sec:selected}

The changes of the normalized symmetric cumulants in the $q_n$ classes can be summarized by asking which fraction of the correlation of $q_n$-integrated events the selection carries. Figures~\ref{fig:decomp1} and~\ref{fig:decomp2} show the normalized symmetric cumulants of $q_n$-integrated events, separated into $\NSC^{\rm sel}$ and $\NSC^{\rm res}$ according to Eq.~\eqref{eq:decomp}. In general, $\NSC^{\rm sel}$ is large only when the selection changes both harmonics of the pair, i.e., when the selected $q_n$ resolves the common driver of their fluctuations.

For the $q_2$ selection in OO collisions, $\NSC^{\rm sel}(3,2)$ is negative and $\NSC^{\rm sel}(4,2)$ is positive, which is consistent with the anticorrelation of $v_3$ and the correlation of $v_4$ with $v_2$ seen in Fig.~\ref{fig:vnq2}. These parts are, however, small: they are at most 8\% of $\NSC(3,2)$ and $\NSC(4,2)$ in the 0--10\% centrality class and decrease towards peripheral collisions. In OO collisions, $\NSC^{\rm res}$ carries more than 90\% of the correlation between $v_2$ and $v_3$, and between $v_2$ and $v_4$, and the same holds for $\NSC(4,3)$ and $\NSC(5,2)$. As one moves to Pb--Pb collisions, $\SC^{\rm sel}$ increases strongly: in the 10--40\% centrality classes, it is 50--71\% of $\SC(3,2)$, 37--42\% of $\SC(4,2)$, 18--34\% of $\SC(4,3)$ and 39--50\% of $\SC(5,2)$, i.e., the $q_2$ selection captures a large fraction of the correlations that involve $v_2$. The exception is $\SC^{\rm sel}(5,3)$, which is small and negative, since the increase of $v_5$ with $q_2$ is accompanied by a decrease of $v_3$. For the $q_3$ selection, $\SC^{\rm sel}(4,2)$ and $\SC^{\rm sel}(5,2)$ are small in both systems, since the $q_3$ classes differ little in $v_2$. In Pb--Pb collisions, where the $q_3$ selection changes $v_3$ strongly, $\SC^{\rm sel}$ is 10--30\% of $\SC(3,2)$, $\SC(4,3)$ and $\SC(5,3)$ in the 10--40\% centrality classes, whereas in OO collisions it is below 1\%. For the $q_2$ selection in Pb--Pb collisions, $\NSC^{\rm sel}(5,4)$ is positive, since both $v_4$ and $v_5$ increase with $q_2$, and it is compensated in part by a negative $\NSC^{\rm res}(5,4)$, i.e., the small $\NSC(5,4)$ of $q_n$-integrated events results from a partial cancellation.

In OO collisions with an acceptance similar to that of ATLAS, the $q_2$ selection thus changes the magnitudes of $v_2$, $v_3$, $v_4$ and $v_5$, but it does not separate the correlations between them. The normalized symmetric cumulants change only moderately, mainly in the lowest $q_2$ class of central collisions, and the selection cannot be used to remove the contribution of $v_2$ to these correlations in the way it partly does in Pb--Pb collisions. Extending the $q_n$ window outward to $|\eta|=6.3$ or 7.8 increases the power of the selection only moderately (Appendix~\ref{app:window}), which implies that these conclusions do not depend on the details of the $q_n$ acceptance.

\section{Summary}
\label{sec:summary}

In summary, this work studies the response of the flow coefficients and the symmetric cumulants to the $q_2$ and $q_3$ selections in OO and Pb--Pb collisions at $\sNN=5.36$~TeV using EPOS4, with an acceptance similar to that of ATLAS. The separation of the symmetric cumulants into $\SC^{\rm sel}$ and $\SC^{\rm res}$ shows directly how much of a correlation the selection controls. It is observed that $\SC^{\rm sel}$ is large only when the selection changes both harmonics of the pair, i.e., when $q_n$ resolves the common driver of their fluctuations. This implies that a change of $\NSC(m,n)$ in a $q_n$ class should be interpreted together with this decomposition. For example, both $q_2$ classes of Pb--Pb collisions give a smaller magnitude of $\NSC(3,2)$, since the selection moves the correlation into the differences between the classes.

The power of the selection is set mainly by the multiplicity in the forward window rather than by the collision system. In OO collisions, the statistical fluctuation of $q_n$ dominates over the flow signal, and $\SC^{\rm res}$ carries more than 90\% of $\SC(3,2)$ and $\SC(4,2)$, whereas it carries about half of them as one moves to mid-central Pb--Pb collisions. Thus, the $q_2$ selection in OO collisions changes the magnitudes of the flow harmonics, but it cannot remove the contribution of $v_2$ to their correlations. Nevertheless, the lowest $q_2$ class of central OO collisions selects events in which $v_4$ is closer to its linear response to $\varepsilon_4$, and the magnitudes of $\NSC(3,2)$ and $\NSC(4,2)$ are 22\% and 29\% smaller in the 0--10\% centrality class. This reduction persists with the 2-subevent method, which hints that it is not caused by nonflow. In Pb--Pb collisions, the $q_2$ and $q_3$ selections act separately on the two factors of the nonlinear term $\chi_{523}V_2V_3$, i.e., a larger $v_2$ shifts the correlation of $v_5$ from $v_2$ to $v_3$, and the $q_3$ selection does the reverse.

EPOS4 does not reproduce the strength of $\NSC(3,2)$ and $\NSC(4,2)$ in central Pb--Pb collisions, and these results therefore serve as a model baseline. However, the mechanisms discussed here follow from the structure of the nonlinear response and from the finite resolution of $q_n$, and are expected to hold beyond this model. This study not only shows that event-shape engineering in light-ion collisions requires a much better resolution of $v_2$ than the forward windows studied here provide, but also establishes event-shape-selected measurements of $\NSC(5,2)$ and $\NSC(5,3)$ in Pb--Pb collisions as a probe of the nonlinear mode coupling.

\begin{acknowledgments}

I gratefully acknowledge support from the Hungarian National Research, Development and Innovation Office (NKFIH) under Contract Nos.~NKFIH NKKP ADVANCED\_25 K153456, 2025-1.1.5-NEMZ\_KI-2025-00005 and 2024-1.2.5-TET-2024-00022. I also acknowledge the Wigner Scientific Computing Laboratory (WSCLAB) in Budapest, Hungary, for providing the computational resources used in this study.

\end{acknowledgments}

\appendix

\section{Comparison with measurements}
\label{app:data}

\begin{figure}
\includegraphics[width=0.9\linewidth]{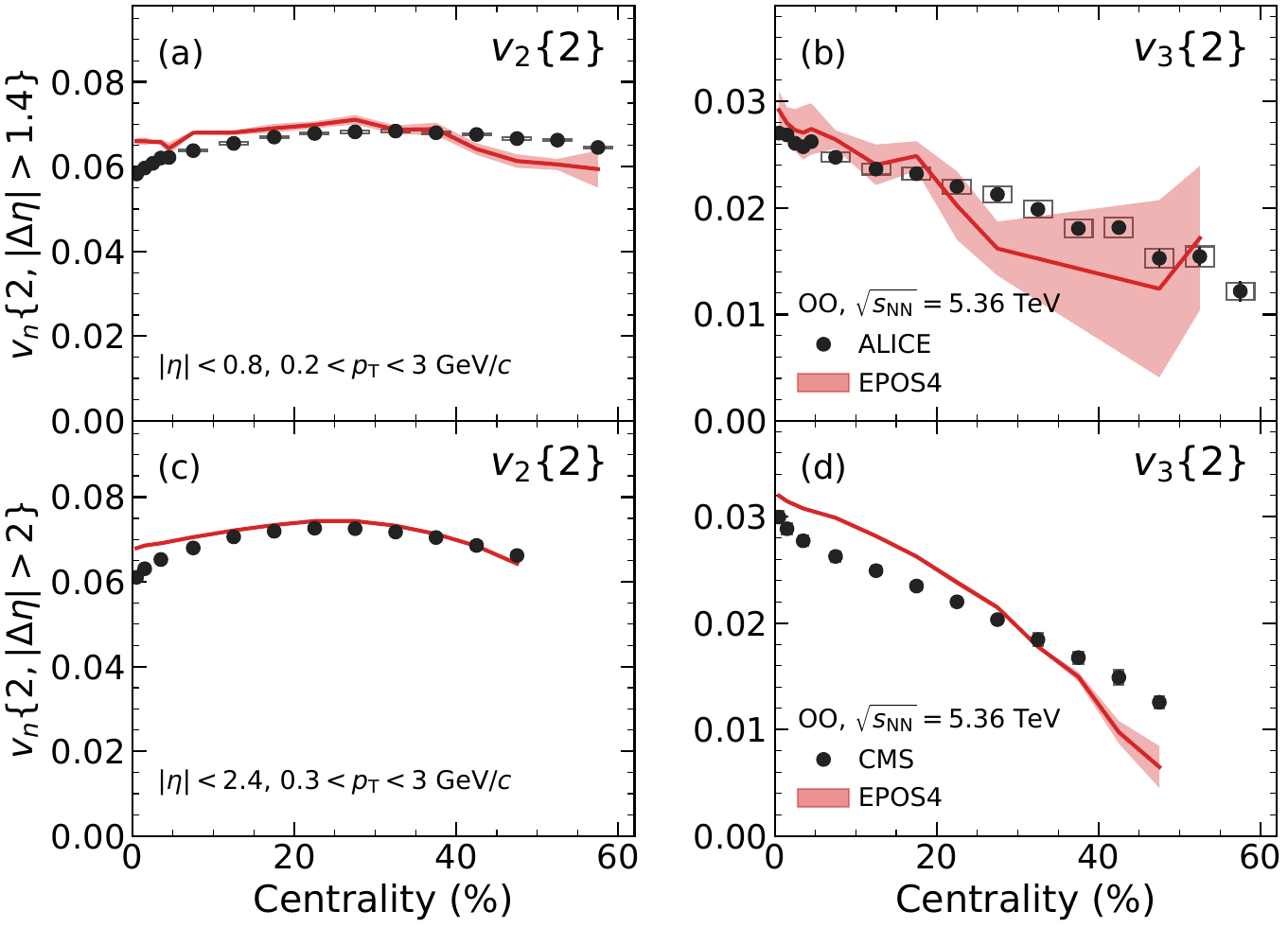}
\caption{$v_2\{2\}$ (left) and $v_3\{2\}$ (right) as a function of centrality in OO collisions at $\sNN=5.36$~TeV measured by ALICE~\cite{ALICE:2025luc} (upper, $|\eta|<0.8$, $0.2<p_{\rm T}<3$~GeV/$c$, $|\Delta\eta|>1.4$) and CMS~\cite{CMS:2025tga} (lower, $|\eta|<2.4$, $0.3<p_{\rm T}<3$~GeV/$c$, $|\Delta\eta|>2$), compared with EPOS4. Bars and boxes show the statistical and systematic uncertainties of the data, and bands the statistical uncertainties of EPOS4.}
\label{fig:dataOO}
\end{figure}

\begin{figure*}[!h]
\includegraphics[width=0.86\textwidth]{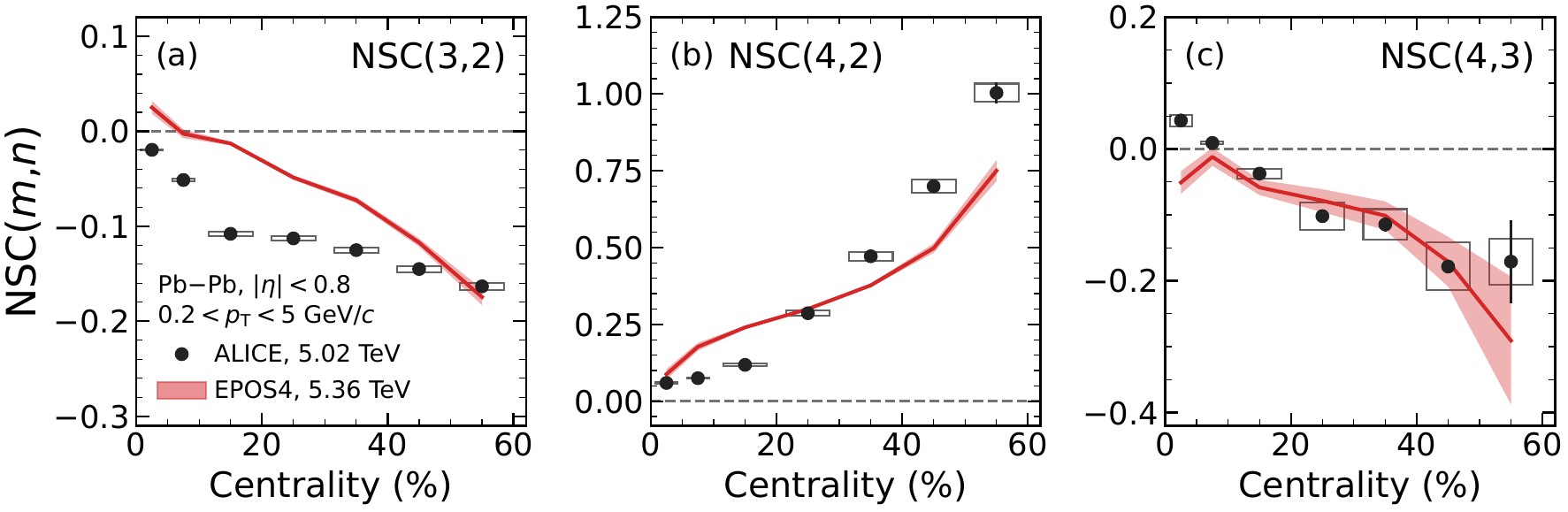}
\caption{$\NSC(3,2)$ (a), $\NSC(4,2)$ (b) and $\NSC(4,3)$ (c) as a function of centrality in Pb--Pb collisions measured by ALICE at $\sNN=5.02$~TeV~\cite{ALICE:2021adw} ($|\eta|<0.8$, $0.2<p_{\rm T}<5$~GeV/$c$), compared with EPOS4 at $\sNN=5.36$~TeV. Bars and boxes show the statistical and systematic uncertainties of the data, and bands the statistical uncertainties of EPOS4.}
\label{fig:dataPbPb}
\end{figure*}

To show how well EPOS4 describes the observables studied here, its results for $q_n$-integrated events are compared with measurements, using the kinematic selection, the pseudorapidity gap and, as far as possible, the centrality estimator of each measurement, i.e., the multiplicity in the FT0C acceptance, $-3.3<\eta<-2.1$, for ALICE in OO collisions~\cite{ALICE:2025luc}, the transverse energy in $3.2<|\eta|<4.9$ for CMS~\cite{CMS:2025tga}, and the multiplicity in the FT0 acceptance for Pb--Pb collisions. As in the measurement~\cite{ALICE:2021adw}, $\NSC(m,n)$ of Pb--Pb collisions uses the four-particle cumulant without a gap and without nonflow subtraction, divided by $\ctwo{m}\ctwo{n}$ of the 2-subevent method with $|\Delta\eta|>1$~\cite{ALICE:2016kpq}. The difference between $\sNN=5.02$ and 5.36~TeV is neglected, since the normalized symmetric cumulants depend only weakly on the collision energy~\cite{ALICE:2017kwu,ALICE:2021adw}.

One can observe in Fig.~\ref{fig:dataOO} that EPOS4 reproduces the weak centrality dependence of $v_2\{2\}$ in OO collisions and agrees with both measurements within about 7\% in the 5--40\% centrality classes, while it is above the data in the most central collisions and below the ALICE values in the 40--60\% centrality classes. For $v_3\{2\}$, EPOS4 lies above the data in central collisions and decreases faster towards peripheral collisions. This hints that EPOS4 has either too small a $v_3$ or too large negative away-side correlations in peripheral OO collisions.

\begin{figure*}[!h]
\includegraphics[width=0.62\textwidth]{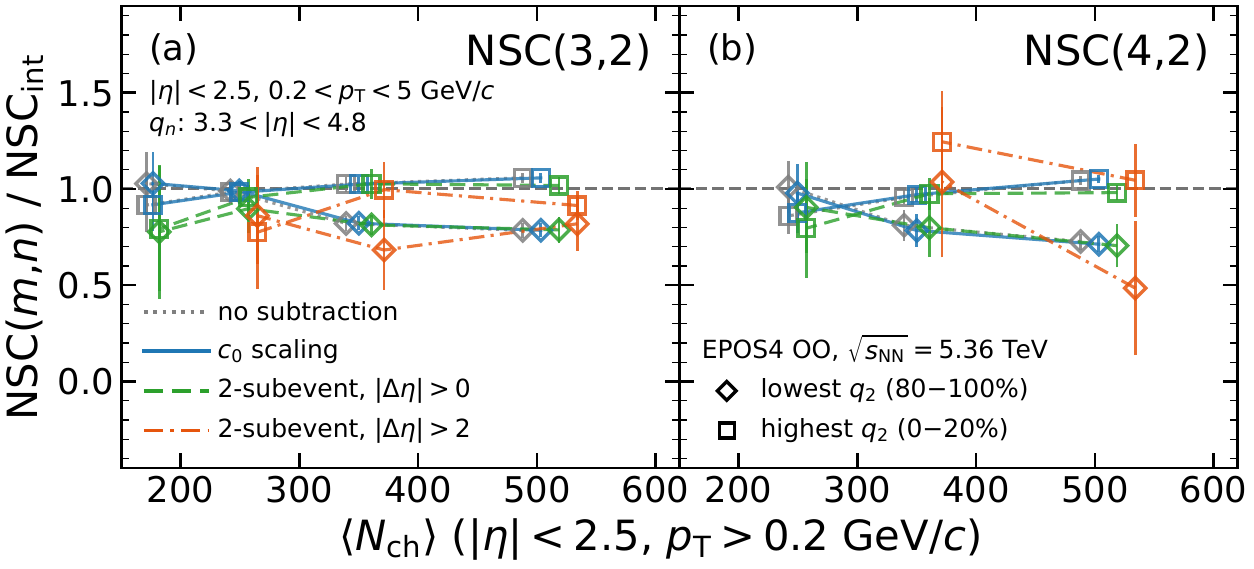}
\caption{Ratio of $\NSC(3,2)$ (a) and $\NSC(4,2)$ (b) in the lowest (diamonds) and highest (squares) $q_2$ classes to $q_n$-integrated events in OO collisions at $\sNN=5.36$~TeV using EPOS4, for four nonflow treatments (line styles). A method is shown only where its $\SC(m,n)$ of $q_n$-integrated events has a significance of at least three standard deviations, and its points are shifted horizontally in the order of the legend. The uncertainties are statistical.}
\label{fig:nonflow}
\end{figure*}

For Pb--Pb collisions (Fig.~\ref{fig:dataPbPb}), the magnitude of $\NSC(3,2)$ in EPOS4 is well below the measured one except in the 50--60\% centrality class, and $\NSC(4,2)$ has a weaker centrality dependence than measured, while $\NSC(4,3)$ agrees with the data in the 10--60\% centrality classes. The results of the main text are therefore a model baseline: the trends with centrality and the signs of the correlators in noncentral collisions are reproduced, but the strength of the correlations in Pb--Pb collisions is not.

\section{Dependence on the nonflow suppression}
\label{app:nonflow}

The reduction of $\NSC(3,2)$ and $\NSC(4,2)$ in the lowest $q_2$ class of central OO collisions (Sec.~\ref{sec:nsc32}) could also arise if the $q_2$ selection changed the nonflow at midrapidity, e.g., through jets that reach both the forward window and the tracks. To test this, the symmetric cumulants are also computed with the 2-subevent method~\cite{Jia:2017hbm,Huo:2017nms}, with two particles from $\eta<-\Delta\eta/2$ and two from $\eta>\Delta\eta/2$, for $\Delta\eta=0$ and~2. This method needs no template, and the denominator of Eq.~\eqref{eq:nsc} is computed with the same subevents.

One can observe in Fig.~\ref{fig:nonflow} that the lowest $q_2$ class is reduced by about the same factor with all methods in the 0--10\% and 10--20\% centrality classes, irrespective of how much nonflow they remove, while the highest $q_2$ class stays close to unity. Without a gap, the reduction has a significance of 3.0 and 3.4 standard deviations for $\NSC(3,2)$ in these two centrality classes, and of 2.6 standard deviations for $\NSC(4,2)$ in the 0--10\% centrality class. With the gap, the uncertainties are four to eight times larger, and the ratios agree with those of the other methods. This implies that the reduction is not caused by nonflow.

\section{Dependence on the $q_n$ window}
\label{app:window}

\begin{figure*}[!h]
\includegraphics[width=0.86\textwidth]{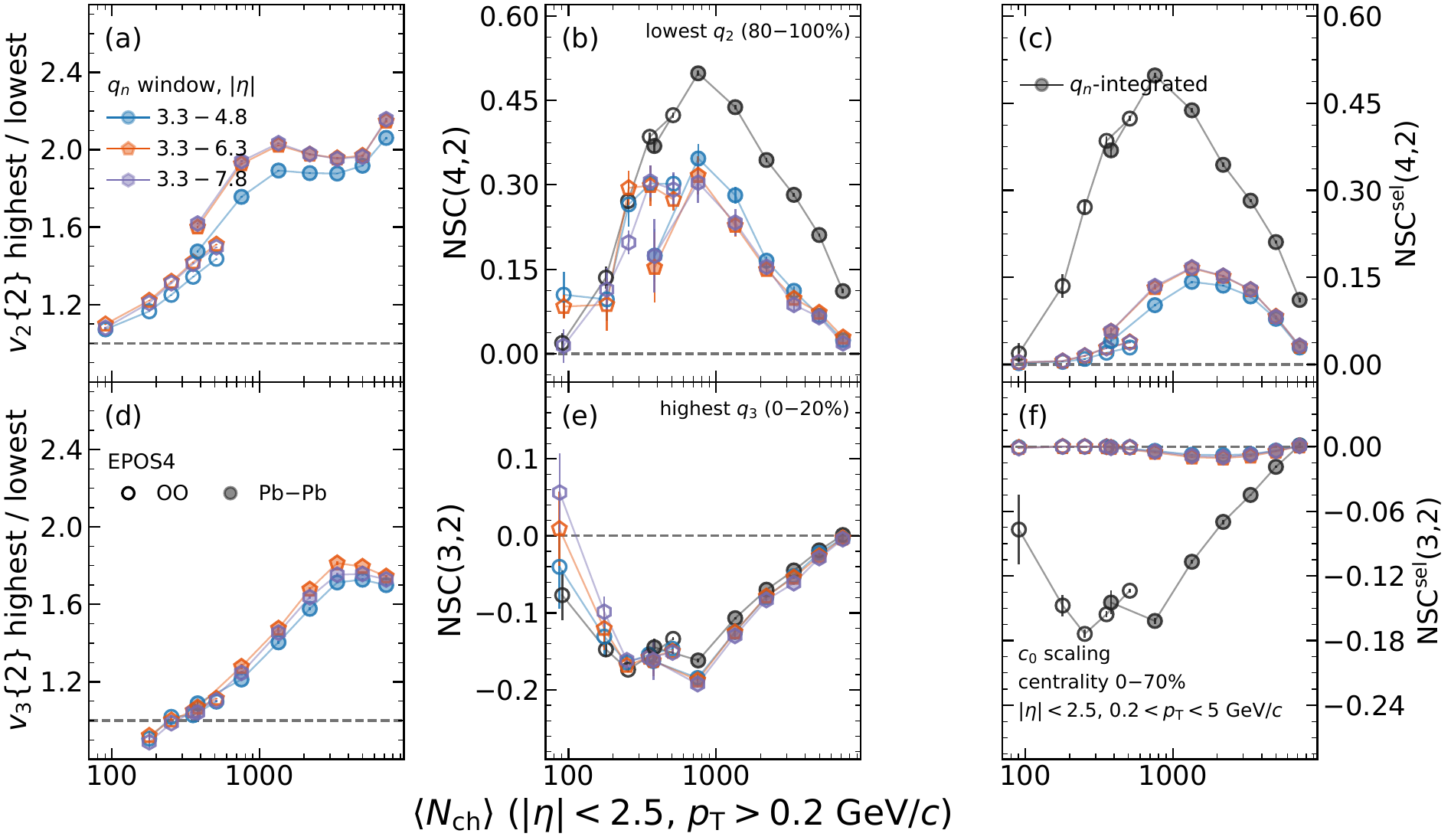}
\caption{Dependence of the $q_2$ (upper) and $q_3$ (lower) selections on the pseudorapidity window of $q_n$, $3.3<|\eta|<4.8$ (default, blue), $3.3<|\eta|<6.3$ (orange) and $3.3<|\eta|<7.8$ (purple), in OO and Pb--Pb collisions at $\sNN=5.36$~TeV using EPOS4, as a function of $\Nch$ for centrality classes 0--70\%. Left: ratio of $\vtwo{n}$ in the highest to that in the lowest $q_n$ class. Middle: $\NSC(4,2)$ of the lowest $q_2$ class (b) and $\NSC(3,2)$ of the highest $q_3$ class (e). Right: $\NSC^{\rm sel}(4,2)$ (c) and $\NSC^{\rm sel}(3,2)$ (f), Eq.~\eqref{eq:decomp}. The values for $q_n$-integrated events are shown in black. The symmetric cumulants use the $\czero$ nonflow subtraction. The uncertainties are statistical.}
\label{fig:window}
\end{figure*}

To test the dependence on the $q_n$ window, $q_n$ is also computed in $3.3<|\eta|<6.3$ and $3.3<|\eta|<7.8$, which extend the default window outward. As shown in Fig.~\ref{fig:window}, the ratio of $\vtwo{2}$ in the highest to that in the lowest $q_2$ class of the 0--10\% OO centrality class increases only from 1.44 to 1.51 with the first window, and the widest window gives no further increase. This is because the correlation with midrapidity decreases with $|\eta|$, so that the region $6.3<|\eta|<7.8$ adds as much fluctuation to $q_n$ as signal.

Accordingly, $\SC^{\rm sel}(3,2)$ and $\SC^{\rm sel}(4,2)$ in the 0--10\% OO centrality class increase only from 7--8\% to 9--11\% of $\SC(3,2)$ and $\SC(4,2)$, and $\NSC(4,2)$ of the lowest $q_2$ class changes by at most 9\% (1.1 standard deviations), so that the $q_n$ window does not change the conclusions.

\clearpage 
\bibliography{refs}

\end{document}